\documentclass[%
 reprint,
 amsmath,amssymb,
 aps,
]{revtex4-1}

\usepackage{graphicx}
\usepackage{dcolumn}
\usepackage{bm}
\usepackage{xcolor}
\usepackage{subfigure}
\usepackage[colorlinks=true,linkcolor=blue,citecolor=blue,urlcolor=blue]{hyperref}

\begin{document}


\author{Sk Noor Alam}
\email{noor1989phyalam@gmail.com}
\affiliation{Aligarh Muslim University, Aligarh, Uttar Pradesh-202002, India}
\author{Victor Roy}
\email{victor@niser.ac.in}
\affiliation{National Institute Of Science Education and Research, Jatni, Odisha-752050, India}
\author{Shakeel Ahmad}
\affiliation{Aligarh Muslim University, Aligarh, UP-202002,India}
\author{Subhasis Chattopadhyay}
\affiliation{Variable Energy Cyclotron Centre, HBNI, Kolkata, WB-700064,India}

\date{\today}
\title{Electro-magnetic field fluctuation and its correlation with the participant plane in Au+Au and isobaric collisions at $\sqrt{s_{NN}}=200$ GeV }

\begin{abstract}
Intense transient electric ({\bf E}) and magnetic ({\bf B}) fields are produced in the high energy heavy-ion collisions. 
The electromagnetic fields produced in such high-energy heavy-ion collisions are proposed to give rise to a multitude
of exciting phenomenon including the Chiral Magnetic Effect. 
We use a Monte Carlo (MC) Glauber model to calculate the electric and magnetic fields, more specifically 
their scalar product $\bf{E}\cdot\bf{B}$, as a function of space-time 
on an event-by-event basis for the Au+Au collisions at $\sqrt{s_{NN}}=200$ GeV for different centrality classes. We also calculate the same for the isobars Ruthenium and Zirconium at $\sqrt{s_{NN}}=200$ GeV. In the QED sector  
$\bf{E}\cdot\bf{B}$ acts as a source of Chiral Separation Effect, Chiral Magnetic Wave, etc., which are associated phenomena to the Chiral Magnetic Effect. We also study the relationships between the electromagnetic symmetry plane angle defined by $\bf{E}\cdot\bf{B}$ ($\psi_{E.B}$) and the participant plane angle $\psi_{P}$ defined from the participating nucleons for the second-fifth order harmonics. 
\end{abstract}

\maketitle

\section{Introduction}
\label{Section:Introduction}




The initial state fluctuations in high-energy heavy-ion collisions play an essential role in understanding several bulk observables. We  can attribute the two primary sources of these initial state fluctuations to the event-by-event~(e-by-e) geometry fluctuations of the nucleon's position inside the nuclei due to the nuclear wave function and the fluctuation in impact strong fluctuating transient electro-magentic~(EM) fields in the overlap zone of the colliding nucleus. 
The EM field generated in high-energy heavy-ion collision experiments such as Relativistic Heavy Ion Collider (RHIC) 
and the Lager Hadron Collider (LHC) is known to be the strongest magnetic field  in the universe (e.g., B $\sim 10^{18}-10^{19}$ Gauss for $\sqrt{s_{NN}}=200$ GeV) \cite{Ref1,Ref2,Ref3,Ref4,Ref5,Ref6}. 
The magnetic field in heavy-ion collisions, while averaged over many events, mostly obey a linear scaling with the centre of mass energy ($\sqrt{s}$) and the impact parameter (b) of collisions, \cite{Ref7} i.e., $\big \langle eB_{y} \big \rangle \sim Zb\sqrt{s}$ for $b\leq 2R_{A}$ where Z is the charge number of the ions, $R_{A}$ is the radius of the nucleus. 
We take the y axis perpendicular to the reaction plane as per the convention, defined 
by the impact parameter(chosen as the x-axis) and the beam direction(z-axis). Furthermore, 
the event-averaged electric fields are also found to be of the same order of magnitude 
as the magnetic fields (e.g., $eB \approx eE \sim 10m_{\pi}^2$ at the topmost RHIC energy Au+Au collisions 
$\sqrt{s_{NN}}=200$ GeV where $m_{\pi}$ is the pion mass). \\ \\
It has been conjectured that in addition to the standard ohmic current driven by the electric field, there might 
appear other new types of current in parity~(P) and charge conjugation ~(C) odd regions in QGP as responses to the electromagnetic fields.
 One of this new type of currents is generated along the background magnetic field, a.k.a. the Chiral Magnetic Effect (CME)\cite{Ref1,Ref8,Ref9,Ref10}. 
In other words, in high-energy heavy-ion collisions, special gluonic configurations (sphalerons and instantons) break the P and the CP in the presence of a strong magnetic field. It results in a global electric charge separation with respect to the reaction plane \cite{Ref11,Ref11a}.
 This charge separation occurs through the transition of the right-handed quarks to the left-handed quarks and vice versa depending on the sign of topological charges \cite{Ref1}. Because of their close association with axial anomaly and the topologically nontrivial vacuum structure of QCD, the CME and other associated phenomena such as chiral separation effect (CSE), the chiral electric separation effect (CESE) is known as anomalous effects~\cite{Huang:2015oca}. 

It is known that only the lowest Landau level contributes to the CME. In the QED sector, combined electric ({\bf E}) and magnetic fields ({\bf B}) are responsible for the transition of chiral fermions from the left-handed chirality branch to the right-handed chirality branch at a rate  $\sim e^2/(2\pi^2) {\bf E} \cdot {\bf B}$ \cite{Ref10,Huang:2015oca}.
 


Similarly, in CSE, the axial current is known to be not conserved due to a source term proportional to $\sim e^3/(2\pi^2) {\bf E} \cdot {\bf B}$.
The same term also appears in the Chiral magnetic wave equation if ${\bf E} \cdot {\bf B}$ is non-zero. In other words,
${\bf E} \cdot {\bf B}$ pumps chirality into the system. In ref.~\cite{Son:2009tf} it was shown that the current conservation equation in a relativistic fluid with one conserved charge, with a $U(1)$ anomaly, contains a source term proportional to $E^{\mu}B_{\mu}$. The scalar product of the four vectors $E^{\mu}$ and $B^{\mu}$ in the fluid rest frame is ${\bf E} \cdot {\bf B}$. 
Hence it is interesting to study ${\bf E} \cdot {\bf B}$ for different collision geometry and its possible correlation with the symmetry (participant) plane, with respect to which we search for the CME signal. It is worthwhile to mention that although the event averaged magnetic field shows a linear behavior with collision centrality, the electric field, on the other hand, shows an opposite trend, i.e., maximum 
for the central collisions and gradually decreases for higher centralities.
In this paper, we focus on the spatial distribution of ${\bf E} \cdot {\bf B}$ for various centrality Au+Au, Ru+Ru, and Zr+Zr collisions at $\sqrt{s_{NN}} = 200$ GeV to investigate their angular correlation with the 
geometry of the fireball.  To this end, we introduce the participant plane $\psi_{EB}$ defined with the weight of ${\bf E} \cdot {\bf B}$, and we show the correlation of it with the participant plane $\psi_{pp}$.

The rest of this paper is organized as follows: In Sec. II, we describe the detail of calculating electromagnetic fields from the Glauber model on an event-by-event basis. We also discuss their impact parameter dependence and event averaged value. In Sec. III. we discuss the main results, which consists of the impact parameter, space-time, and system size dependence of ${\bf E} \cdot {\bf B}$ and its correlation with the participant plane. Finally, we summarise this study in Sec. IV.

\section{Calculation of electric and magnetic field}
\label{Section:EB Field}
Customarily the electromagnetic field generated by a relativistic charged particle is calculated from the 
well known Liénard–Wiechert potentials, however, we will calculate it from the second-rank antisymmetric 
electromagnetic field tensor $F^{\alpha\beta}=\partial^{\alpha} A^{\beta}-\partial^{\beta} A^{\alpha}$ using
the Lorentz transformation. Here $A^{\mu}$ is the four-potential
due to an electric charge, in the following calculations we assume the charged protons inside 
the colliding nuclei move in a straight line trajectory and there are neglegible change in momentum 
after the collision. The calculation goes as follow \cite{Ref12}: first we calculate the component of electromagnetic 
fields and corresponding $F^{\prime \gamma\delta}$ in the rest frame $S^{\prime}$ of the charge particle.
The fields in the laboratory frame is calculated from $F^{\alpha\beta}$ which is obtained from $F^{\prime\gamma\delta}$
through the Lorentz transformation:

\begin{equation}
F^{\alpha \beta}=\frac{\partial x^{\alpha}}{\partial x^{\prime\gamma}} \frac{\partial x^{ \beta}}{\partial x^{\prime\delta}} F^{\prime\gamma \delta} ,
\end{equation}
or in matrix notation $F=\Lambda F^{\prime} \tilde{\Lambda}$, where $\Lambda$ is the matrix representation 
of the Lorentz transformation
and $\tilde{x^{\mu}}$ corresponds transpose of $x^{\mu}$. We choose a boost $\beta=v_{z}$ along the $z$ axis.
In this case it can be easily shown that the electric fields transform as
\begin{eqnarray}
E_{x}&=&\gamma E_{x}^{\prime}+\gamma \beta B_{y}^{\prime}, \\
E_{y}&=&\gamma E_{y}^{\prime}-\gamma \beta B_{x}^{\prime}, \\
E_{z}&=&E_{z}^{\prime},
\end{eqnarray}
and the magnetic fields transform as,
\begin{eqnarray}
B_{x}&=&\gamma B_{x}^{\prime}-\gamma \beta E_{y}^{\prime}, \\
B_{y}&=&\gamma B_{y}^{\prime}+\gamma \beta E_{x}^{\prime}, \\
B_{z}&=&B_{z}^{\prime}.
\end{eqnarray}
Since the charge is at rest in the $S^{\prime}$ frame $B_{x}^{\prime}=B_{y}^{\prime}=B_{z}^{\prime}=0$,
furthermore, it is easy to verify ${\bf B}={\bf \beta} \times {\bf E}$. Next, we calculate $ {\bf E}^{\prime}$ at a point 
P $(x,y,z)$ at time $t$ for a charge at ($x^{\prime}_{c},y^{\prime}_{c},z^{\prime}_{c}$) at time $t^{\prime}$ by noting that 
$z^{\prime}_{c} \approx \beta t^{\prime}$ (we assume that the origin of the lab frame $S$ and the moving frame $S^{\prime}$ coincide at $t=t^{\prime}=0$). The subscript $c$ corresponds to the charge.
For convenience, we denote the transverse distance $\zeta=\sqrt{(x^{\prime}_{c}-x)^2+(y^{\prime}_{c}-y)^2}$, the distance 
from the charge to $P$ is $r^{\prime}=\sqrt{\zeta^2+\beta^2 t^{\prime 2}}$ (here we have taken the center of the nucleus to be at the 
origin of $S^{\prime}$). Also, the positions of the charged particles in the transverse plane are 
assumed to be frozen due to large Lorentz $\gamma=(1-\beta^2)^{-1/2}$. A straightforward calculation gives the following values of the electric
fields in $S$ frame
\begin{eqnarray}
\label{eq:Ex}
E_{x}&=&\frac{\gamma q x}{\left(\zeta^{2}+ \beta^{2} \gamma^{2}\left(t-\beta z\right)^{2}\right)^{3 / 2}}, \\
\label{eq:Ey}
E_{y}&=&\frac{\gamma q y}{\left(\zeta^{2}+ \beta^{2} \gamma^{2}\left(t-\beta z\right)^{2}\right)^{3 / 2}}, \\
\label{eq:Ez}
E_{z}&=&\frac{q \beta \gamma (t-\beta z) }{\left(\zeta^{2}+ \beta^{2} \gamma^{2}\left(t-\beta z\right)^{2}\right)^{3 / 2}},
\end{eqnarray}
and the magnetic fields are given by 
\begin{eqnarray}
\label{eq:Bx}
B_{x}&=&\frac{-\gamma \beta q y}{\left(\zeta^{2}+ \beta^{2} \gamma^{2}\left(t-\beta z\right)^{2}\right)^{3 / 2}}, \\
\label{eq:By}
B_{y}&=&\frac{\gamma \beta q x}{\left(\zeta^{2}+ \beta^{2} \gamma^{2}\left(t-\beta z\right)^{2}\right)^{3 / 2}}, \\
\label{eq:Bz}
B_{z}&=&0.
\end{eqnarray}


The total electromagnetic field at any point is evaluated using the principle of superposition i.e.,
calculating fields using Eq.\eqref{eq:Ex} - \eqref{eq:Bz} for all the protons inside the nucleus. 
We use a cutoff value of $\zeta$ = 0.3 fm while calculating the electric and magnetic field usig Eq.\eqref{eq:Ex} - \eqref{eq:Bz} 
We note that this cutoff value was chosen as 
an average effective distance between the quarks inside the nucleons, and it was also reported \cite{Ref6} that there is a 
weak dependence of the field values on $\zeta$ in the range 0.3 to 0.6 fm.
Since the colliding nucleus at $\sqrt{s_{NN}}=200$ GeV has Lorentz $\gamma \sim 100$, we can safely assume the nucleus as a flat disk that has a vanishing thickness along the $z$ axis. Also, due to the time-dilation, the nucleons will appear as frozen inside the nucleus, and all nucleons effectively move along $z$ with constant $v_z$ i.e 
${\bf v_n} \equiv $ (0,0,$v_z$). As per the convention, we take the velocity of the target nucleus as $+v_z$, and the velocity of the projectile nucleus is $-v_z$. $v_z$ is calculated from the ratio of the relativistic momentum and the energy of a proton 
\begin{equation}
v_z = \sqrt{1-\Big(\frac{2m_p}{\sqrt{s_{NN}}}\Big)^2}.
\label{Eq:vz}
\end{equation}

To obtain the nucleon positions 
we use the MC-Glauber model \cite{Ref13}. We also calculate the initial spatial eccetricity ($\epsilon$, defined later)
and the number of participating nucleons $(N_{\rm{part}})$ for a given impact parameter from the MC-Glauber model. 
In the MC-Glauber model, the positions of the nucleons inside the nucleus are determined by the nuclear density function measured in low-energy electron scattering experiments\cite{Ref14}. The functional form of this distribution is:
\begin{equation}
\rho(r,\theta) = \frac{\rho_0}{1+exp\big[\frac{r - R\big(1+ \beta_2 Y_2^{0}(\theta) + \beta_4 Y_4^{0}(\theta) \big)}{a}  \big]},
\label{Eq:rho_rtheta}
\end{equation}
where $\rho_0$ corresponds to the nuclear density at the center, $R$ is the radius of the nucleus, 
$a$ is the skin depth (it controls how quickly the nuclear density falls off near the edge of the nucleus). The spherical harmonics $Y_l^{m}(\theta)$ and parameters $\beta_2$ and $\beta_4$ are used to measure the deformation from spherical shape. 
For our study, we take $R = 6.38$ fm, $a = 0.535$ fm, and $\beta_2=\beta_4=0$ for the $Au_{79}^{197}$ nucleus. We use parameter $\beta_2^{Ru}$ =0.158 , $\beta_2^{Zr}$=0.08, $R^{Ru}=5.085$ fm, $R^{Zr}=5.02$fm and a = 0.46 fm for both Ru and Zr \cite{Ref14a,Ref14b,Ref14c}. $\beta_4$ is taken 0 for both nuclei. 

\begin{figure}[htb]
\centering
\includegraphics[width=.40\textwidth]{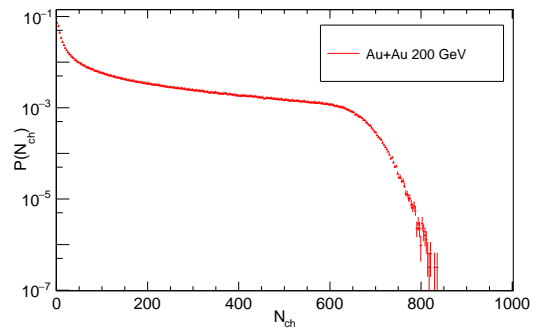}
\caption{(Color online) Probability distribution of the charged particles multiplicity in Au+Au $\sqrt{s_{NN}} = 200$ GeV collisions.}
\label{fig:mult}
\end{figure}

We sample the nucleon positions assuming that they are randomly distributed with the given distribution $4\pi r^2 \rho(r)$ (integrating on $\theta$ and $\phi$ ) for Au nucleus and $r^2\sin(\theta) \rho(r,\theta)$ for Ru and Zr nuclei respectively. \\
The impact parameters $b$ of the collisions are randomly selected from the distribution $\frac{dN}{db} \sim b$ upto a maximum 
value of $\simeq 20$ fm $>2R$. The center of the target and projectile nuclei are shifted 
to $(-\frac{b}{2},0,0)$ and $(\frac{b}{2},0,0)$ respectively. We use the inelastic nucleon-nucleon cross section 
$\sigma_{NN} = 42 $ mb for the top RHIC energy $\sqrt{s_{NN}} = 200 $ GeV for calculating the probability of an 
interaction between the target and the projectile nucleons\cite{Ref15,Ref16}.
To show the centrality dependence, we calculate the centrality of the collisions using $N_{\rm{part}}$ 
and the number of binary collisions $(N_{\rm{coll}})$ obtained from the MC Glauber model . The multiplicity for a given 
$N_{\rm{part}}$  and $(N_{\rm{coll}})$ is calculated using the two component model as
\begin{equation}
\frac{dN_{ch}}{d\eta} = n_{\rm{pp}}\left[ (1-x)\frac{N_{\rm{part}}}{2} + xN_{\rm{coll}}\right],
\label{Eq:mult}
\end{equation}
where $x$ is the fraction of hard scattering, $n_{\rm{pp}}$ is the average multiplicity  
per unit pseudo-rapidity in pp collisions.
The above two-component model of the particle production is based on the assumption that 
the average particles produced through the soft interactions are proportional to the $N_{\rm{part}}$,
and the probability of hard interactions is proportional to $N_{\rm{coll}}$.
We calculate the centrality of a given collision in the following way:
the number of independent particle emitting sources for a given impact parameter 
are $(1-x)\frac{N_{\rm{part}}}{2} + xN_{\rm{coll}}$. Each of these sources produces particles following a negative 
binomial distribution (NBD) with a mean $\mu$ and the width $\sim$ 1/k,
\begin{equation}
\textit{P}_{\mu,k}(n) = \frac{\Gamma(n+k)}{\Gamma (n+1) \Gamma (k)} \Big(\frac{\mu}{\mu+k}\Big)^n \Big(\frac{k}{\mu+k}\Big)^k.
\label{Eq:NBD}
\end{equation}
 $\textit{P}_{\mu,k}(n)$ is the probability of measuring n hits per independent sources.
The mean of this NBD distribution is calculated from the pseudo-rapidity density of 
the charged multiplicity for the non-single diffractive $\bar{p}p$ collisions at a given 
$\sqrt{s_{NN}}$ energy \cite{Ref17}:
\begin{equation}
\mu = \mathcal{A}\ln^2{(s_{NN})} - \mathcal{B}\ln{(s_{NN})} + \mathcal{C},
\label{Eq:nn}
\end{equation}
where $\mathcal{A}=0.023\pm0.008$, $ \mathcal{B}=0.25\pm0.19$, and $\mathcal{C}=2.5\pm1.0$.
The charged particle multiplicity data for Au+Au 200 GeV collisions measured by the 
STAR collaboration are explained for $x = 0.13$ and $k = 1.7$, and $\mu = 2.08$ for 
the pseudo-rapidity range i.e$\mid \eta \mid < 0.5$ \cite{Ref18}.
For example, we show the charged particle multiplicity distribution for Au+Au collisions 
at $\sqrt{s_{NN}} = 200$ GeV for $|\eta | < 0.5$ in Fig.~\ref{fig:mult}.
To calculate the centrality of collisions, we subdivide the total area in Fig.~\ref{fig:mult} into different bins with the condition that
the fractional area corresponds to a particular centrality.
For example, the bin boundaries $n40$ and $n50$ for the  40$\%$-50$\%$ centrality are defined in such a way 
that the following relation holds:
$\frac{\int_{\infty}^{n40} P(N_{ch}) dN_{ch}}{\int_{\infty}^{0} P(N_{ch}) dN_{ch}} = 0.4$,  and $\frac{\int_{\infty}^{n50} P(N_{ch}) dN_{ch}}{\int_{\infty}^{0} P(N_{ch}) dN_{ch}} = 0.5$.  \\\\ 
We use three centrality bins 0$\%$-5$\%$, 40$\%$-50$\%$ and 70$\%$-80$\%$ for our calculation of the electric and magnetic field; corresponding impact parameter ranges are $0-3.2$, $9.3-10.6$, and $12.2-13.5$ fm, which are very similar to the values given in \cite{Ref18}. 

\begin{figure}[htb]
\centering
\includegraphics[width=.50\textwidth]{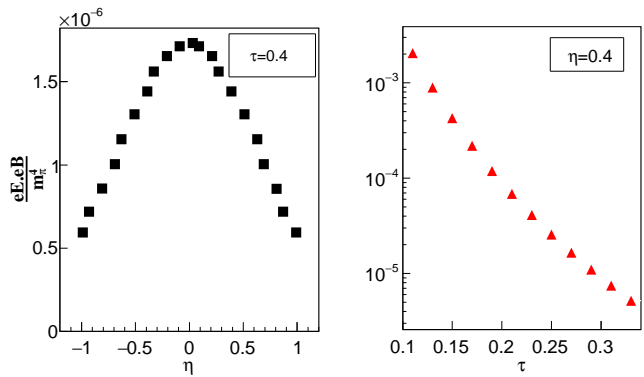}
\caption{(Color online) Left panel: event averaged distribution of $\bf{E}\cdot\bf{B}$ for 40$\%$-50$\%$ centrality Au+Au collisions at $\sqrt{s_{NN}} = 200$ GeV as a function of $\eta$ at constant proper time $\tau$=0.4. Rght panel: same as the left panel but as a function of $\tau$ at constant $\eta$=0.4.}
\label{fig:milneedotb40To50Cent}
\end{figure}

As mentioned earlier, the topology of the electromagnetic fields in heavy-ion collisions has non-trivial dependence on the centrality (possibly also on the $\sqrt{s_{\rm{NN}}}$). Consequently, the source ($\bf{E}\cdot\bf{B}$) of the chiral current in the transverse plane also has a non-trivial centrality dependence. The axial current generated by the magnetic field is supposed to predominantly flow along the direction perpendicular the participant plane. Hence, we investigate here how the sources $\bf{E}\cdot\bf{B}$ of this current is correlated to the 
participant plane. From the perspective of heavy-ion collisions it is customary to use the Milne co-ordinates i.e., we use the longitudinal proper time $\tau=\sqrt{t^2-z^2}$ ,$x$,$y$, and the space-time rapidity $\eta = \frac{1}{2} \log(\frac{t+z}{t-z})$ instead of the Cartesian co-ordinate.  In Fig.\ref{fig:milneedotb40To50Cent} we show the distribution of 
${\bf E}\cdot {\bf B}$ for 40$\%$-50$\%$ centrality with $\eta$ (left plot) and $\tau$ (right plot). In left plot, we calculate $\bf{E}\cdot\bf{B}$  
in the forward light-cone spanned by the region for $\tau$ = 0.4  and $-1.0\leq \eta \leq 1.0$. In right plot, we also calculate $\bf{E}\cdot\bf{B}$  
in the forward light-cone. But for this plot,  phase space is spanned by the region for $\eta$ = 0.4  and $0.1\leq \tau \leq 0.34$.
While calculating the dot product, we transform the field components  $B_{x(y)}$ from the Cartesian to the 
Milne coordinate $\tilde{B}_{x(y)}$ by the following expression
$\tilde{B}_{x} = B_{x}/\cosh\eta, \tilde{B}_{y} = B_{y}/\cosh\eta$ and $\tilde{B}_{z} = B_{z}/\tau$.
The components of the electric field transform similarly.

To investigate the distribution of $\bf{E}\cdot\bf{B}$ (from now on denoted as $\mathcal{E}$) to the participant plane, we introduce the $\bf{E}\cdot\bf{B}$ symmetry plane
 $\psi_{\mathcal{E}}$  defined as \cite{Ref19,Ref20}
\begin{equation}
\epsilon_{n}e^{in\psi_{\mathcal{E}}} = -\frac{\int dxdy r^{n} e^{in\varphi} \mathcal{E}(x,y)}{\int dxdy r^{n} \mathcal{E}(x,y)},
\label{Eq:psi}
\end{equation}
where $r^2 = (x-<x>)^2 + (y-<y>)^2$ and $tan(\varphi) = \frac{y-<y>}{x-<x>}$. Here $(<x>,<y>)$ corresponds to the mean position of the participating nucleons. Using these definitions and Eq.\eqref{Eq:psi} we obtain $\psi_{\mathcal{E}}$ as

\begin{equation}
\psi_{\mathcal{E}} = \frac{1}{n} arctan \frac{\int dxdy r^{n} \sin(n\varphi) \mathcal{E}(x,y)}{\int dxdy r^{n} \cos(n\varphi)\mathcal{E}(x,y)} +\frac{\pi}{n}.
\label{Eq:psi1}
\end{equation}
Before going into the main results of this paper, let us very briefly
go through the impact parameter dependence of the electric and magnetic field produced 
in Au+Au collisions at $\sqrt{s_{\rm{NN}}}=200$ GeV. These results are not new and have 
already been reported in several works \cite{Ref6,Ref7,Huang:2015oca,Ref20ab}, but we include them for the sake of completeness.

\subsection{Impact parameter dependence of the field}
\label{Section: b dependence}
\begin{figure}[htb]
\centering
\includegraphics[width=.40\textwidth]{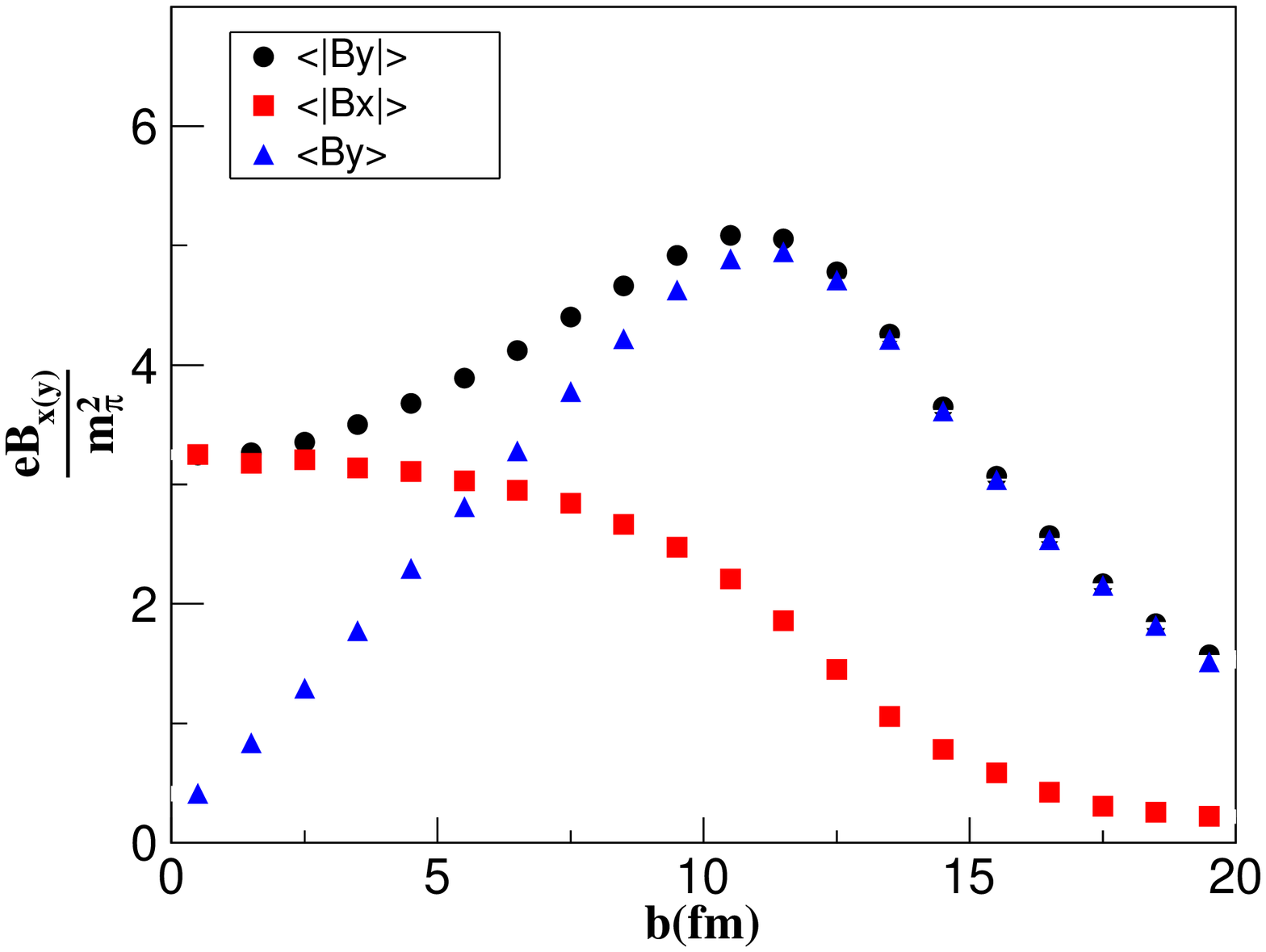}
\includegraphics[width=.40\textwidth]{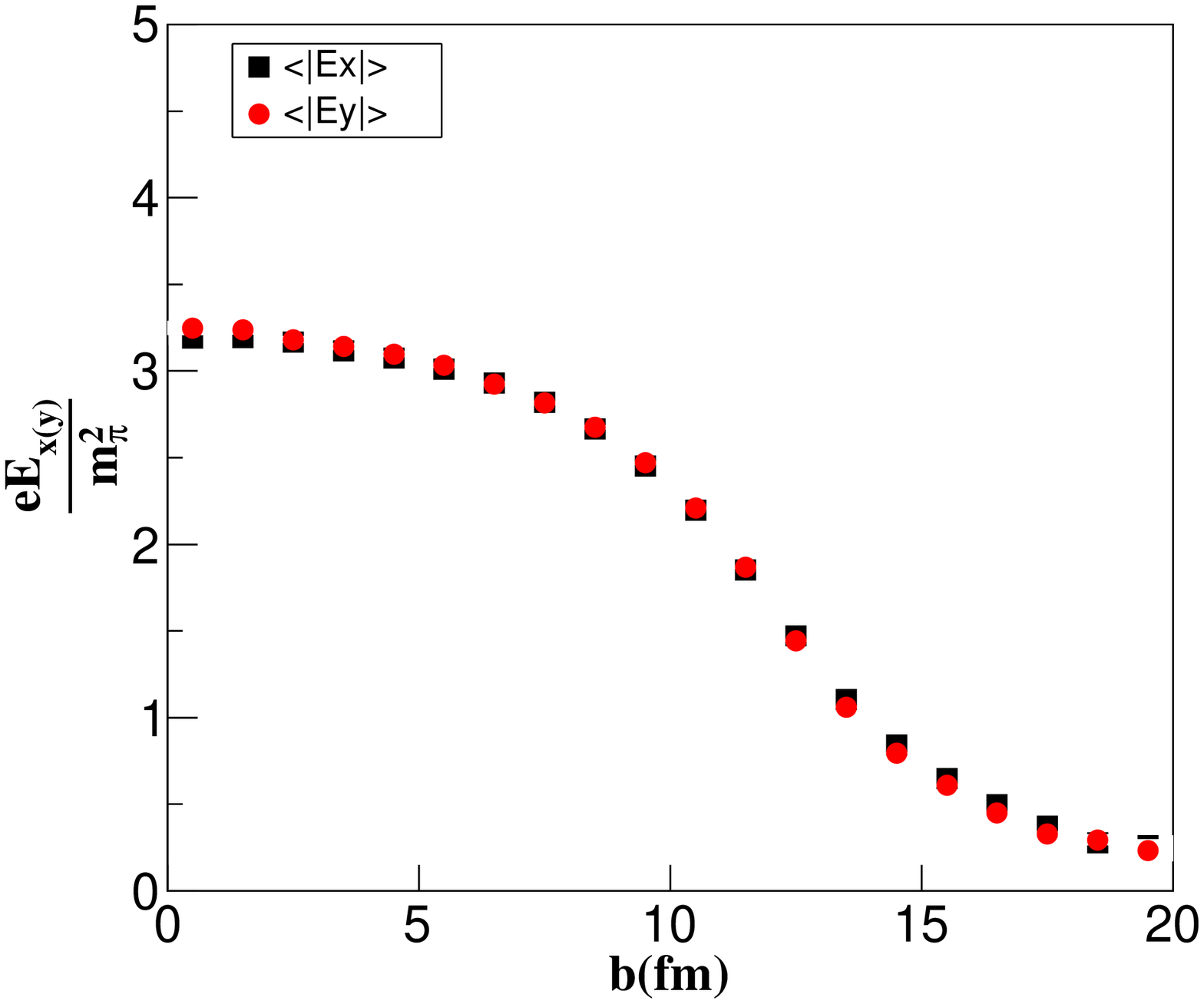}
\caption{(Color online) Top panel: event averaged absolute value of $B_{x}$ and $B_{y}$ (in unit of $m_{\pi}^{2}$) at $\textbf{r}=0$ and t=0  vs impact parameter for Au+Au collisions at $\sqrt{s_{NN}} = 200$ GeV. Bottom panel: same as top panel but for the electric fields.}
\label{fig:modBwithImpact}
\end{figure}

We calculate fields on a regular space-time grid and consider one million events while calculating event-averaged
quantities. The magnitude of the electric and magnetic fields may become
very large near the charges due to the obvious $r$ dependence of Coulomb law, relativistic enhancement of field,
and due to the clustering of charges due to the quantum fluctuations of nuclear wavefunction \cite{Ref20a}. 
But these fluctuations smoothes out when taking event-average
of the field due to its vector nature. These fluctuations might play an essential role in CME; however, here, we show the event-averaged values of electromagnetic fields as a function of the impact parameter of the collisions.
The impact parameter dependence of electric and magnetic field at the origin, i.e., ($x$=0, $y$=0 in our grid space) at $t$=0 is shown in Fig.\ref{fig:modBwithImpact}. 
 Because of the symmetry in the system considered here, 
 $\langle E_{x} \rangle$ and $\langle E_{y} \rangle$ are zero at the origin. It is also interesting to note that $\langle \mid E_{x} \mid \rangle \approx \langle \mid E_{y} \mid \rangle \approx \langle \mid B_{x} \mid \rangle$
 (figure Fig.\ref{fig:modBwithImpact}). 
 From these two figures, we also notice that the electric field decreases as the impact parameter increases while the magnetic field follows the opposite trend. Our results seems to be consistent with \cite{Ref6,Ref7}.

\begin{figure}[htb]
\centering
\includegraphics[width=.40\textwidth]{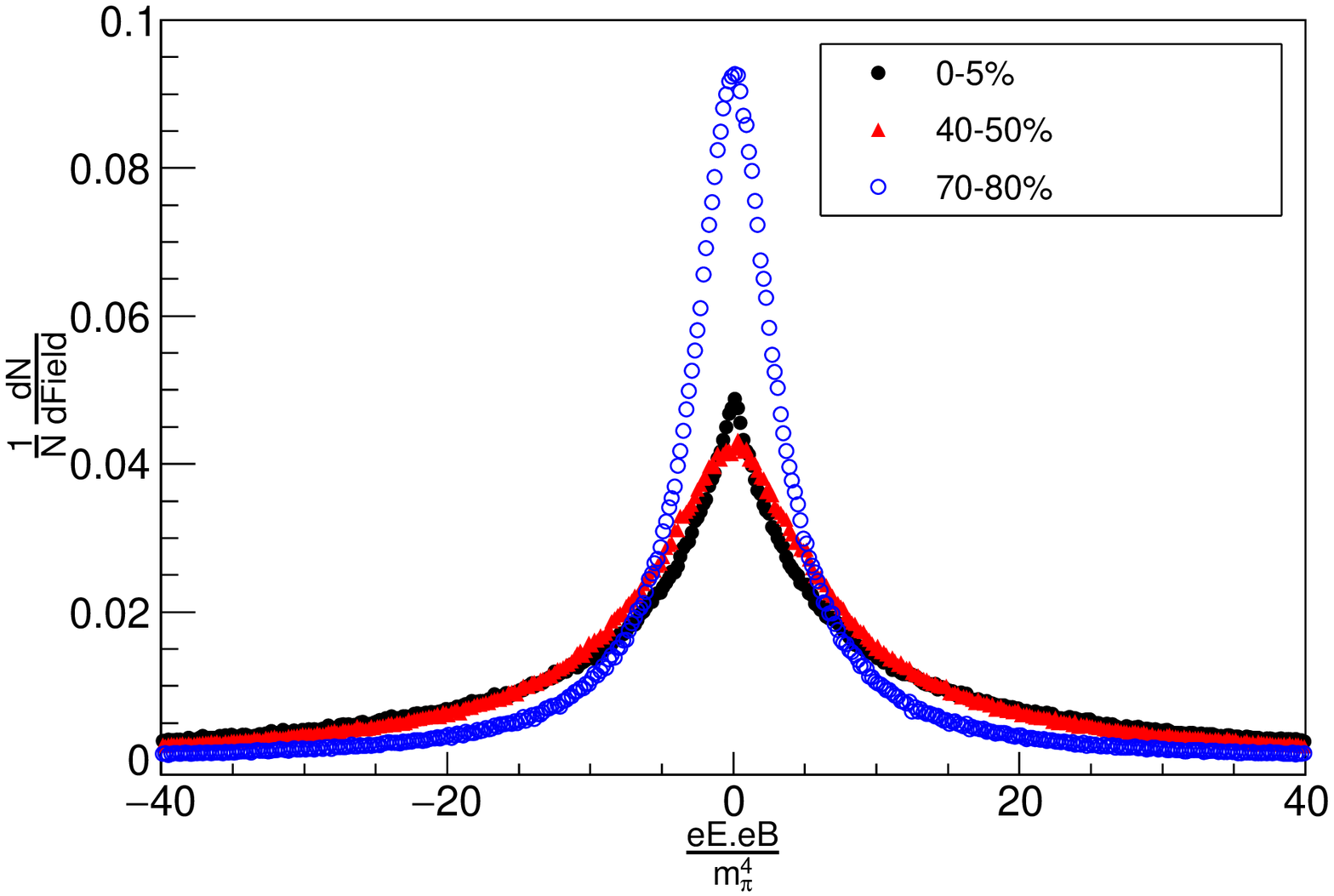}
\includegraphics[width=.40\textwidth]{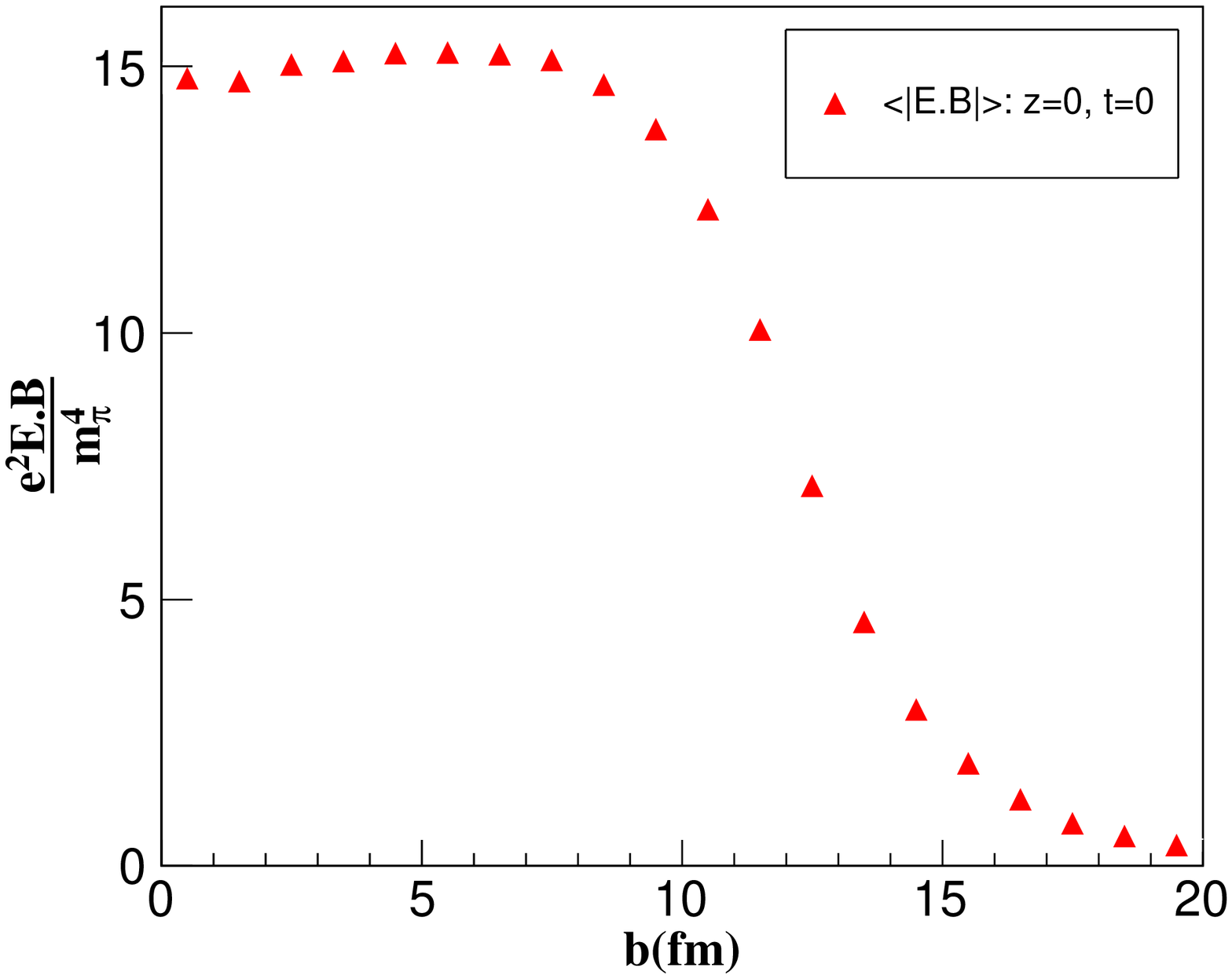}
\caption{(Color online) (top panel) histogram of $\bf{E}\cdot\bf{B}$ at $\textbf{r}=0$ and t=0 for 0-5$\%$, 40-50$\%$ and 70-80$\%$ centrality of Au+Au collisions at $\sqrt{s_{NN}} = 200$ GeV. (bottom panel) Event averaged absolute value of $\bf{E}\cdot\bf{B}$ at $x=y=z=0$ and t=0 fm  as a function of the impact parameter for Au+Au collisions at $\sqrt{s_{NN}} = 200$ GeV.}
\label{fig:histFieldEdotB}
\end{figure}
We end this section with this brief discussion; let us turn to the quantity of our interest in the next section.

\section{Results and discussion}
\label{Section: reusltsA}
\subsection{ impact parameter dependence of $\bf{E}\cdot\bf{B}$}

\begin{figure}[htb]
\centering
\includegraphics[width=.40\textwidth]{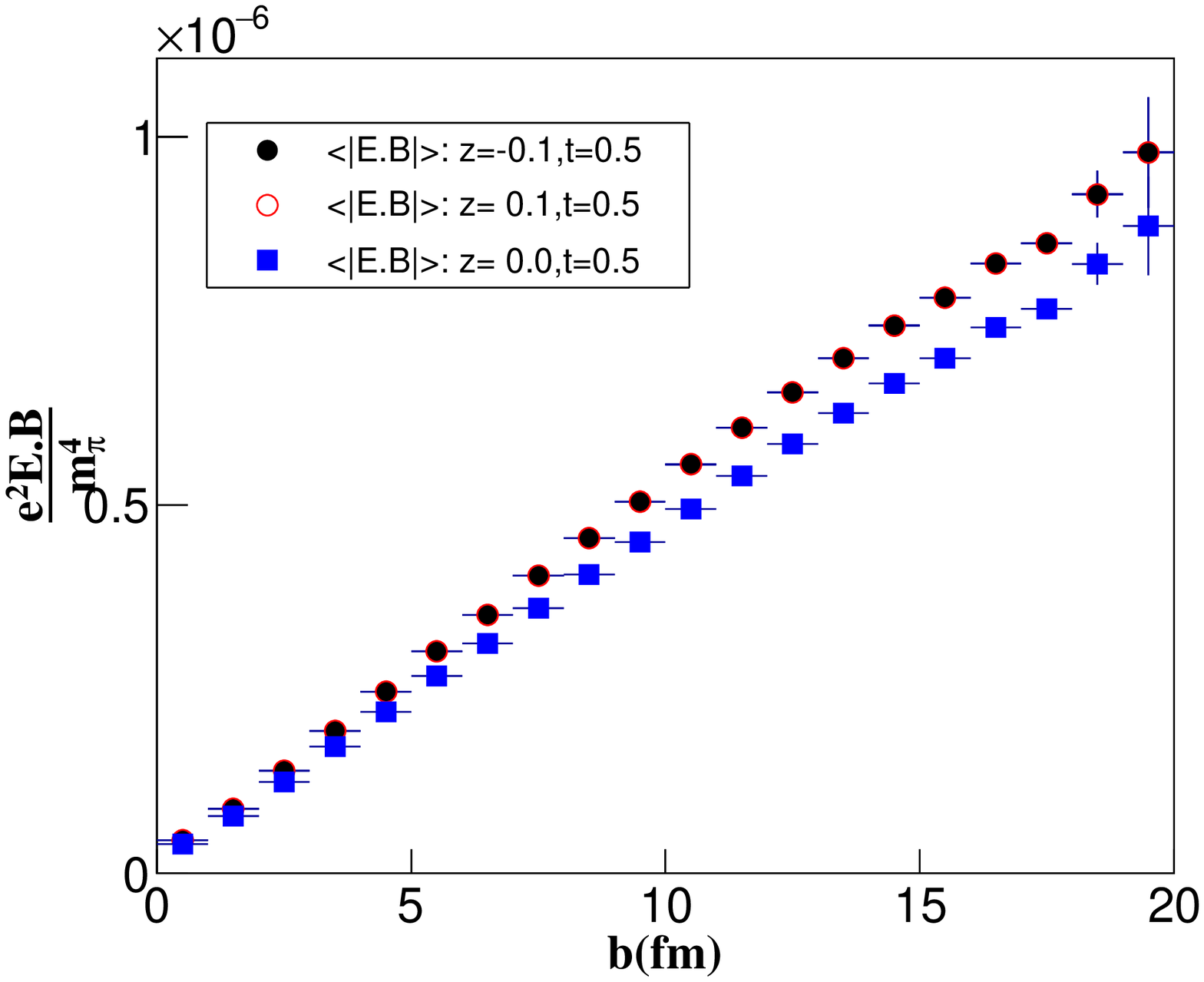}
\includegraphics[width=.40\textwidth]{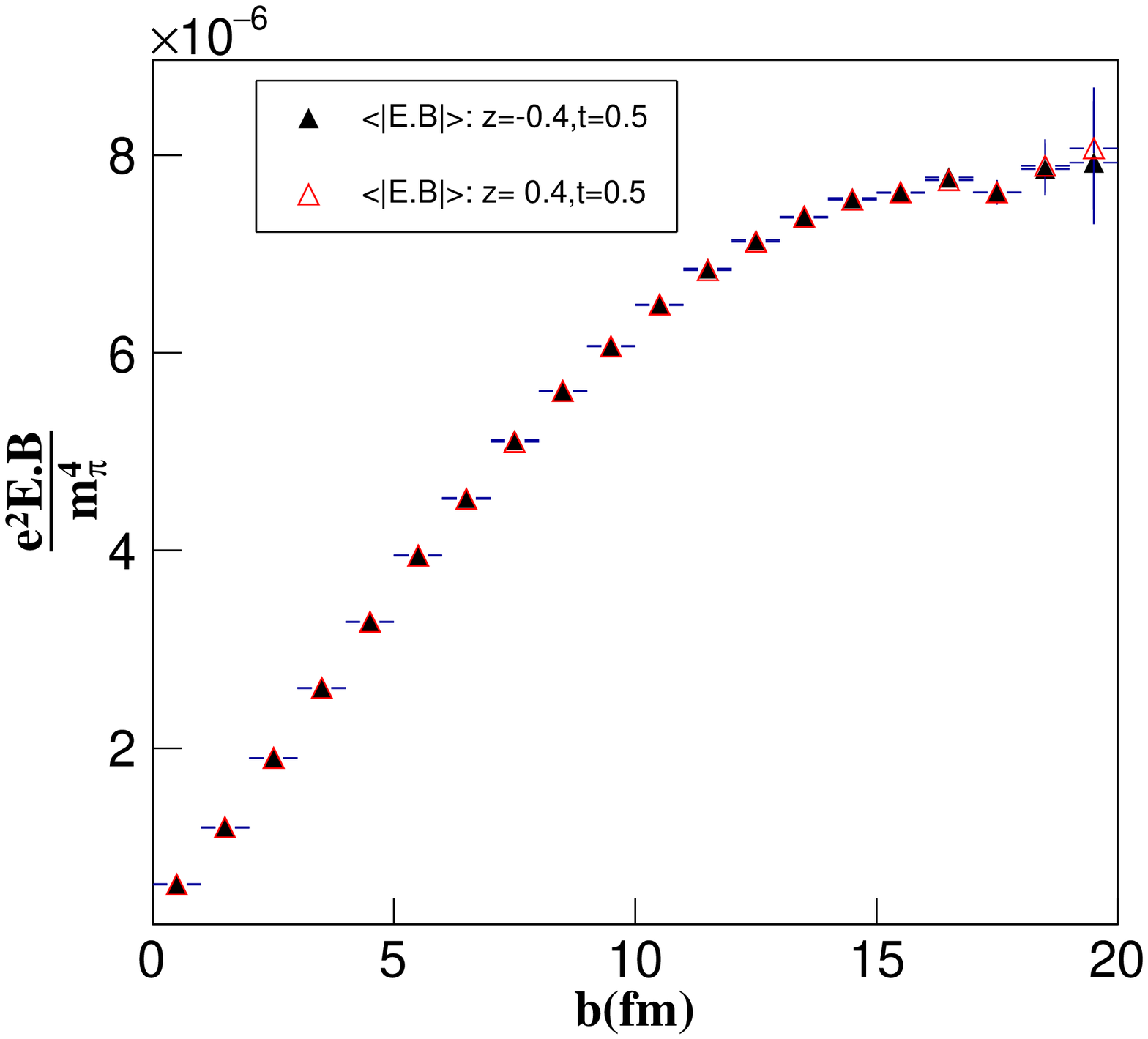}
\caption{(Color online) Event averaged absolute values of $\bf{E}\cdot\bf{B}$ at $x=y=0$  for $|z|=0.1$,$z=0$ (top panel) and $|z|=0.4$ fm (bottom panel) with the impact parameter for Au+Au collisions at $\sqrt{s_{NN}} = 200$ GeV. All results are for t=0.5 fm.}
\label{fig:cprofileEB1}
\end{figure}

In Fig.\ref{fig:histFieldEdotB}, we show one 
dimensional histogram of $\mathcal{E}(0,0)$ distribution for $z=t=0$  and for 
three different centralities 0$\%$-5$\%$ (black circles), 
40$\%$-50$\%$ (red triangles),  and 70$\%$-80$\%$ (open blue circles) Au+Au
 collisions at $\sqrt{s_{NN}} = 200$ GeV. First, we note that there is a non-monotonic 
 dependence of $\mathcal{E}$ on the collision centrality. For peripheral collisions 
 maximum events have $\mathcal{E}\sim 0$, and the distribution also becomes 
 narrower compared to the central/mid-central collisions. This behaviour can be understood 
 as a consequence of near vanishing electric fields at the centre of the collision zone in peripheral 
 collisions at $t=0$ and at midrapidity. 
 It is clear from top panel of Fig.\ref{fig:histFieldEdotB} that although the mean of $\mathcal{E}$ (at the origin) is zero the variance is non-zero. Hence it is more relevant to study the absolute value of the event averaged $\mathcal{E}$. In the bottom panel of Fig.\ref{fig:histFieldEdotB}, and in both panels of Fig.\ref{fig:cprofileEB1} we show the event averaged absolute values of $\mathcal{E}(0,0)$ for $t=z=0$ fm, and $t$=0.5 fm, $z=\pm0.1$,$z=\pm0.4$
respectively. The important difference between these two results is that for the first case
($t=z=0$ fm) electromagnetic fields from the two colliding nuclei almost equally contributed in $\mathcal{E}$, whereas, for the other case, the fields due to each nucleus will be significant for $|z|\sim t$ and is dominated by the nearest nucleus. 
From the bottom panel of Fig.\ref{fig:histFieldEdotB} we observe that for $t=z=0$ fm $\mathcal{E}$ is almost flat up to impact parameter $b\sim10$ fm, and after that, it falls rapidly.
 This observed impact parameter dependence of $\mathcal{E}$ can be attributed to the fact that the magnetic field increases with impact parameter, whereas electric fields diminish.
 The magnitude of $\mathcal{E}$ (expressed in the unit of pion mass) is comparable to the corresponding magnetic fields in central collisions. It may be an over-optimistic claim at this stage; however, this large values of $\mathcal{E}$ at mid-rapidity at the initial time possibly indicates that the CME signal may have a significant contribution from $\mathcal{E}$ along with $B$. To get the complete picture, we must wait for the late time behavior of $\mathcal{E}$ discussed next.
 For the other case, we consider fields at later time $t=0.5$ fm, and at forward and backward regions, i.e., $z\neq 0$. The top panel of Fig.\ref{fig:cprofileEB1} shows the dependence of $\mathcal{E}$ as a function of $b$ for $z$=0 (blue squares), $z=\pm 0.1$ (open and filled circles). It is not surprising that $\mathcal{E}$ for this case is approximately six orders of magnitude smaller than $t=z=0$ case, this is because the electromagnetic fields decay rapidly after the collision, and in this case, only one nucleus significantly contributes to the fields. 
 
From the top panel of Fig.\ref{fig:cprofileEB1} we also observe that $\mathcal{E}$ increases almost linearly with the impact parameter; this is almost opposite to what we observe for $t=z=0$. The impact parameter dependence of $\mathcal{E}$ at finite $z$ is, however, non-trivial, as can be seen from the bottom panel of Fig.\ref{fig:cprofileEB1}, where the same result is shown but for $z=\pm 0.4$. Here one notices that $\mathcal{E}$ rises almost linearly for small $b $ ($<10$ fm), but after that, it starts saturating. Since $z=\pm 0.4$ at time $0.5$ fm is nearer to the receding nuclei, we get a larger value of $\mathcal{E}$ compared to $z$=0 , and $z=\pm 0.1$ (top panel). To conclude this section, we note that at peripheral collisions, charge separation is experimentally observed to be larger than central collisions \cite{Ref21, Ref22}, and it might be linked to the observed $b$ dependence $\mathcal{E}$, along with the magnetic fields, which are also most prominent in mid-central/peripheral collisions.


\subsection{$\bf{E}\cdot\bf{B}$ correlation with participant plane}
In \cite{Ref23}, the correlation of the fluctuating magnetic field with the participant plane showed that a sizable suppression of the angular correlations exists between the magnetic field and the second and fourth harmonic participant planes were found in very central and very peripheral collisions. The importance of space averaged $e^2 {\bf E} \cdot {\bf B}$ was studied recently in \cite{Ref24}  as a function of time for 200 GeV Au+Au collisions at $t= 0.08$ fm and for b= 9 fm. We notice that our finding of the spatial distribution of $\mathcal{E}$ in Au+Au collisions is similar to \cite{Ref24}. In this section, we further investigate the spatial distribution $\mathcal{E}$ and its correlation with the participant plane $\psi_{p}$ by calculating the symmetry plane defined in Eq.\eqref{Eq:psi1}. $\psi_{p}$ is calculated from Eq.\eqref{Eq:psi1} using the positions of wounded nucleons, and by setting $\mathcal{E}=1$ which gives the usual definition used in literature.

Since the isobaric collisions of Ru+Ru and Zr+Zr are important for searching the CME signal, we include results for these two nuclei along with the Au+Au collisions discussed in the previous section. 

\begin{figure}[htb]
\centering
\includegraphics[width=.40\textwidth]{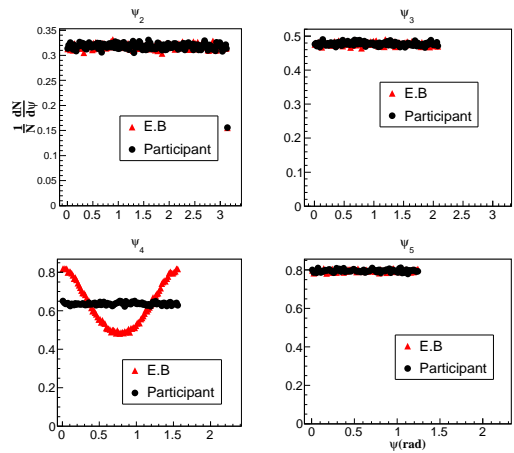}
\caption{(Color online) distribution of $\psi_{\mathcal{E}}^{n}$ and $\psi_{P}^{n}$ ($n$=2-5) for  impact parameter b = 0 fm Au+Au collisions at $\sqrt{s_{NN}} = 200$ GeV.}
\label{fig:corrPsiEBb0}
\end{figure}

\begin{figure}[htb]
\centering
\includegraphics[width=.40\textwidth]{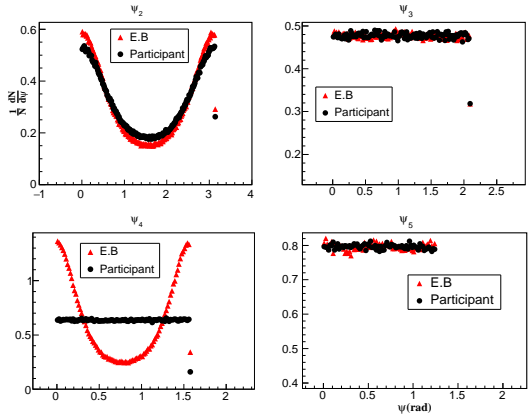}
\caption{(Color online) distribution of $\psi_{\mathcal{E}}^{n}$ and $\psi_{P}^{n}$ ($n$=2-5) for 0$\%$-5$\%$ centrality Au+Au collisions at $\sqrt{s_{NN}} = 200$ GeV.}
\label{fig:corrPsiEB0}
\end{figure}

Let us first discuss the result for Au+Au collisions. We show the distribution of $\psi_{\mathcal{E}}$ and $\psi_{P}$ at $z$=0 
for time $t$=0 fm in Fig.\ref{fig:corrPsiEBb0} for perfectly head on ($b$=0 fm) Au+Au collisions at $\sqrt{s_{NN}} = 200$ GeV.
Since a head-on collision creates an almost symmetric overlap zone, the existence of a particular symmetry plane due to the
participants ($\psi_{p}$) may be ruled out in this case. That is what we observe here from Fig.\ref{fig:corrPsiEBb0}, where black circles show $\psi_{P}^{2}$; the distribution is almost flat. The distribution of $\psi_{\mathcal{E}}^{2}$ is also similar to $\psi_{P}^{2}$; the third and fifth-order symmetry plane show a similar trend.  Interestingly, the rotational symmetry in head-on collisions seems to be broken for $\psi_{\mathcal{E}}^{4}$. The reason behind this behavior, however, is unclear to us. 
We know that the probability of occurrence of a perfectly head-on collision is approximately zero. 
Hence, next in Fig.\ref{fig:corrPsiEB0} we show results for 0$\%$-5$\%$ centrality Au+Au collisions at $t=z=0$ fm.

\begin{figure}[htb]
\centering
\includegraphics[width=.40\textwidth]{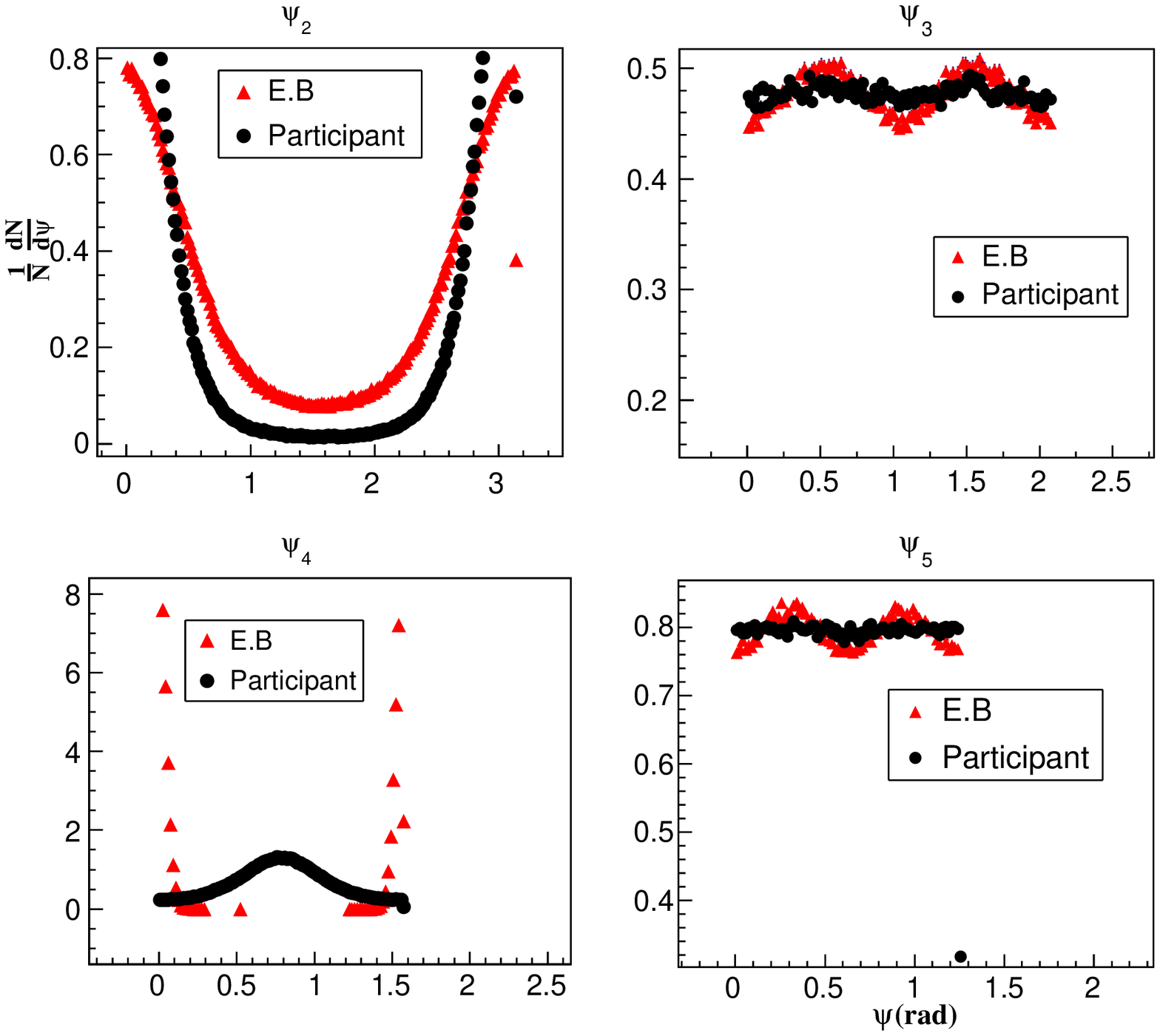}
\caption{(Color online) distribution of $\psi_{\mathcal{E}}^{n}$ and $\psi_{P}^{n}$ ($n$=2-5) for 40$\%$-50$\%$ centrality Au+Au collisions at $\sqrt{s_{NN}} = 200$ GeV.}
\label{fig:corrPsiEB4050}
\end{figure}

\begin{figure}[htb]
\centering
\includegraphics[width=.40\textwidth]{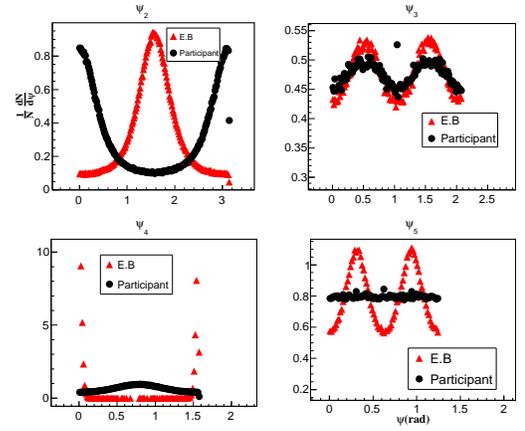}
\caption{(Color online) distribution of $\psi_{\mathcal{E}}^{n}$ and $\psi_{P}^{n}$ ($n$=2-5) for 70$\%$-80$\%$ centrality Au+Au collisions at $\sqrt{s_{NN}} = 200$ GeV.}
\label{fig:corrPsiEB2}
\end{figure}

\begin{figure}[htb]
\centering
\includegraphics[width=.40\textwidth]{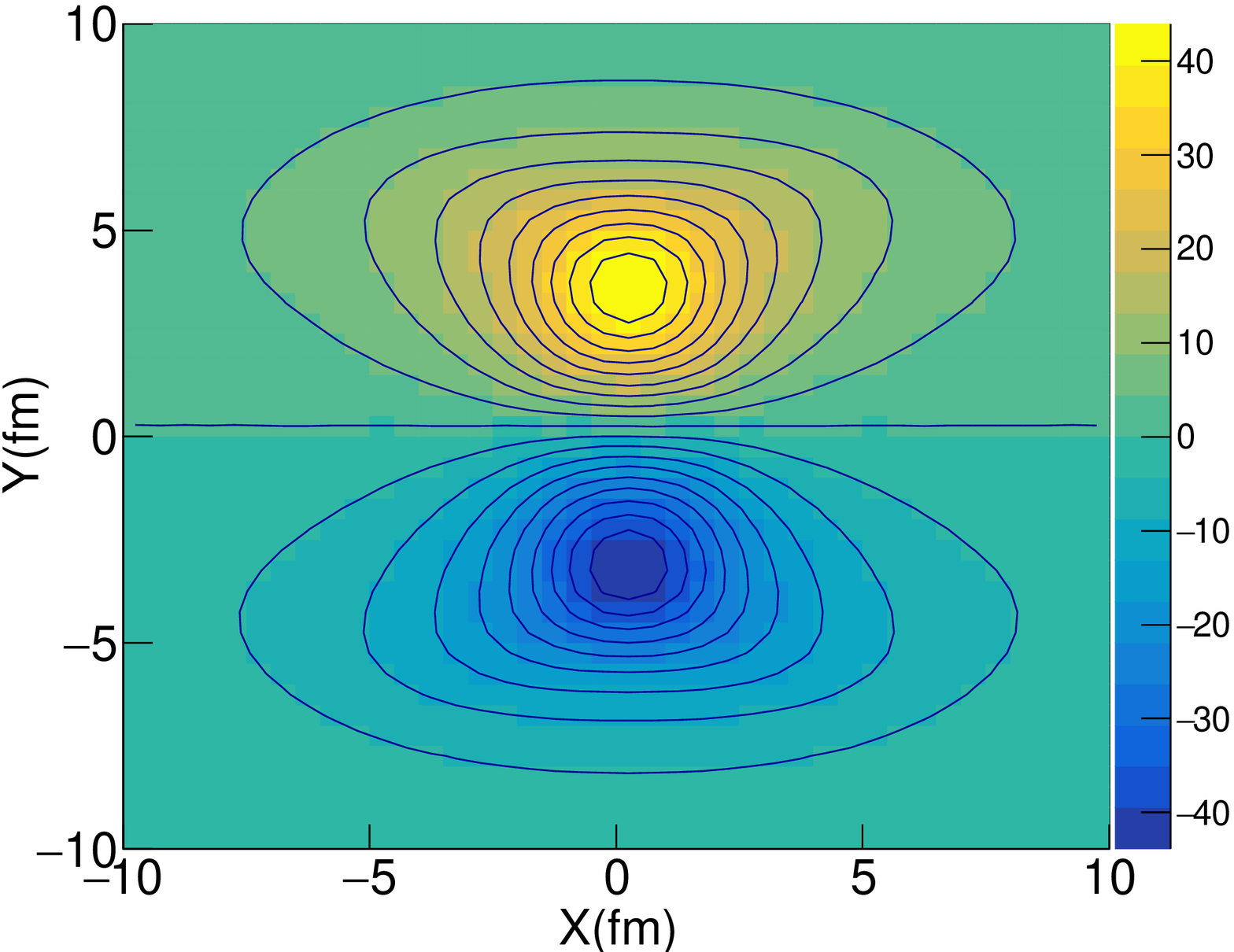}
\includegraphics[width=.40\textwidth]{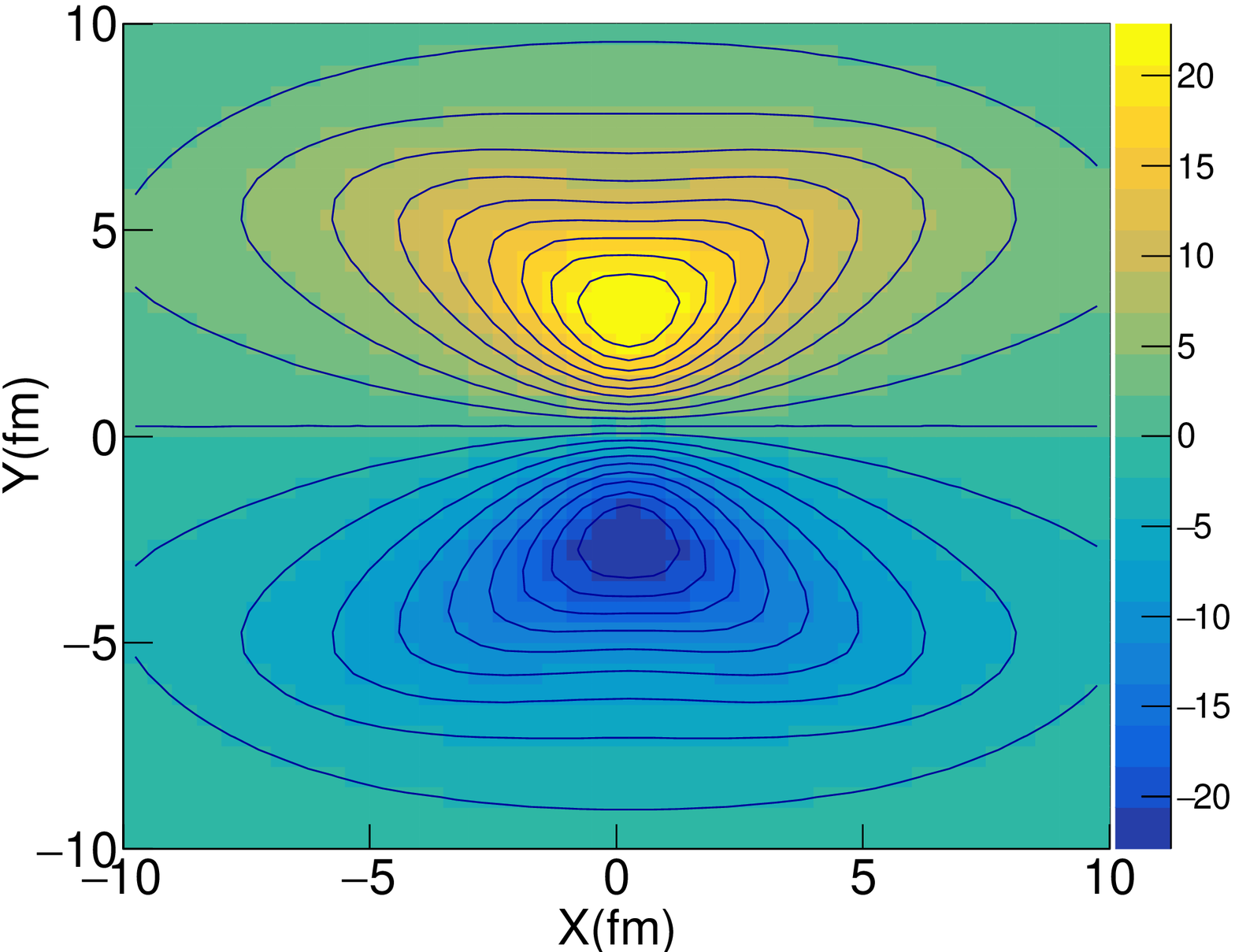}
\caption{(Color online) Top panel: spatial distribution of $\mathcal{E}$ at $z=t=0$ for 40$\%$ -50$\%$ centrality Au+Au collisions at $\sqrt{s_{NN}} = 200$ GeV. Bottom panel: same as the top panel but for 70$\%$ -80$\%$ centrality.}
\label{fig:edotb40To50Cent}
\end{figure}

As expected, in contrast to the $b=0$ case, in Fig.\ref{fig:corrPsiEB0} we see that due to the overlap geometry and the fluctuating nucleon positions, the distribution of the second-order participant plane (black circles) reflects the broken rotational symmetry of the collision zone.
 $\psi_{\mathcal{E}}^{2}$ (red triangles) due to the electromagnetic fields seems to be highly correlated with $\psi_{P}^{2}$.  Other 
 higher order $\psi_{P}$'s in central collisions known to be fluctuating widely and the same is observed here as well. Notably, $\psi_{\mathcal{E}}^{4}$ shows a different trend than $\psi_{P}^{4}$, this is because in central collisions inside the fireball the resultant electric fields
 due to the target and the projectile are much smaller than the magnetic fields; the magnetic fields have a dipole nature, and the 
 corresponding symmetry plane almost coincides with $\psi_{P}^{2}$. This can be more clearly seen from Fig. \ref{fig:corrPsiEB4050} 
 for 40$\%$-50$\%$ centrality collisions,  where the electric fields become vanishingly small, and the magnetic fields are larger, in that case 
  $\psi_{\mathcal{E}}^{4}$ (bottom panel) becomes more oriented along $\psi_{P}^{2}$ (top panel). If we further increase the collision 
  centrality and consider 70$\%$-80$\%$ collisions (see Fig. \ref{fig:corrPsiEB2}) we observe a noticeable change in $\psi_{\mathcal{E}}^{2}$
  as compared to the mid-central collisions. 
  It is clear that the distribution of $\mathcal{E}$ has a $\pi/2$ rotation compared to the central 
  collisions.  To better understand this rotation of symmetry plane for peripheral collisions we 
   we show the contours of $\mathcal{E}$ in the transverse plane for the 40$\%$-50$\%$ (top panel), and  70$\%$-80$\%$ (bottom panel) 
  centralities at $z=0$ and $t =0$ fm in Fig.\ref{fig:edotb40To50Cent}. We can see that a quadrupole like structure appears for 
  70$\%$-80$\%$  collisions.  From the above discussion, we can conclude that $\mathcal{E}$ distribution in the transverse plane for Au+Au collisions at 200 GeV per nucleon is highly correlated with the geometry of the fireball.



Ruthenium and Zirconium nuclei carry the same number of nucleons (96), but Ru has 44 protons, and Zr has 40 protons.
In other words, they have similar shapes and sizes but different electrical charges, which implies different electromagnetic fields generated in Ru+Ru and Zr+Zr collisions. This feature makes them interesting systems for detecting CME by eliminating possible backgrounds.
For the following results we use a fixed impact parameter collisions to keep things simple. 

\begin{figure}[htb]
\centering
\includegraphics[width=.40\textwidth]{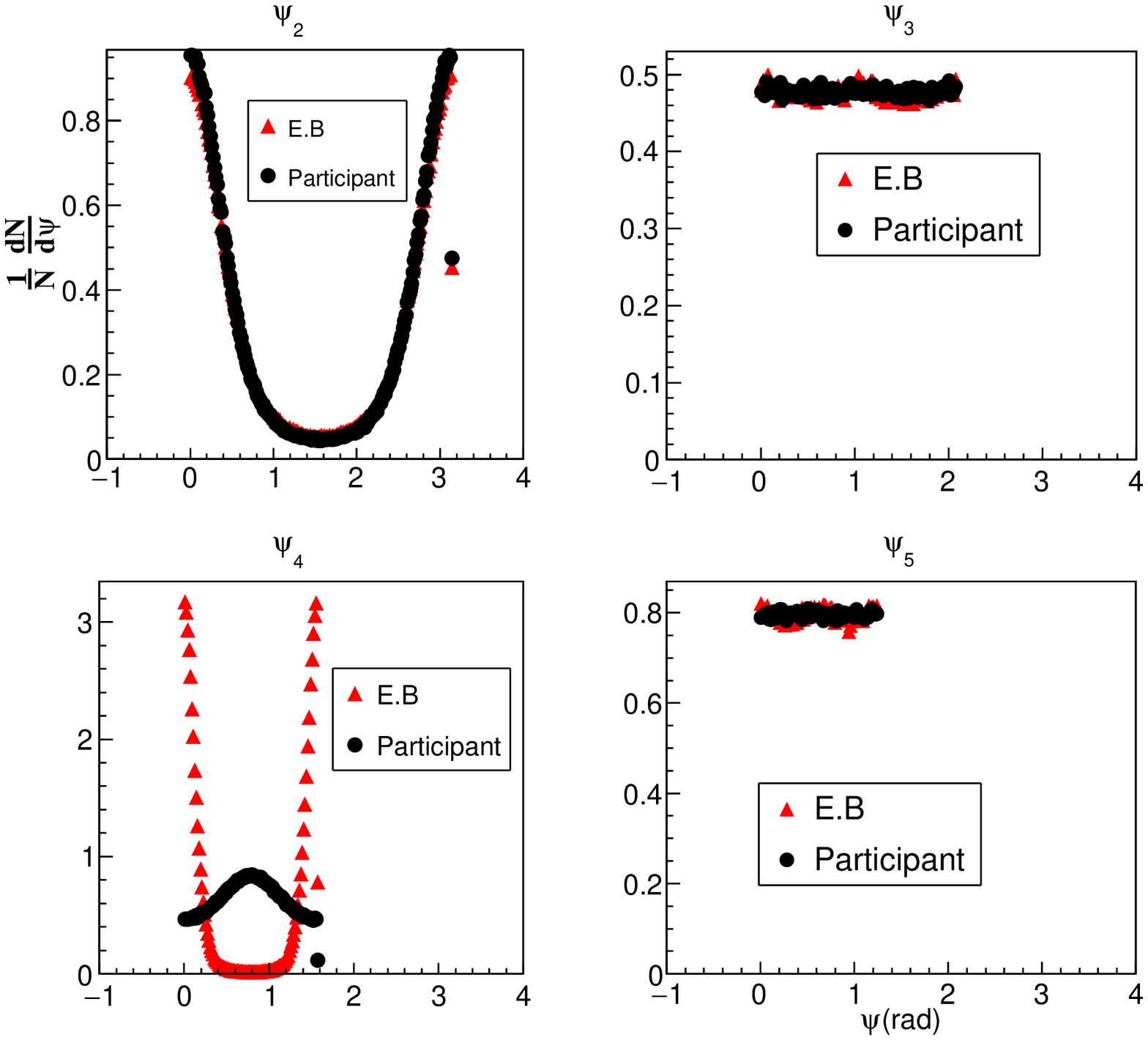}
\includegraphics[width=.40\textwidth]{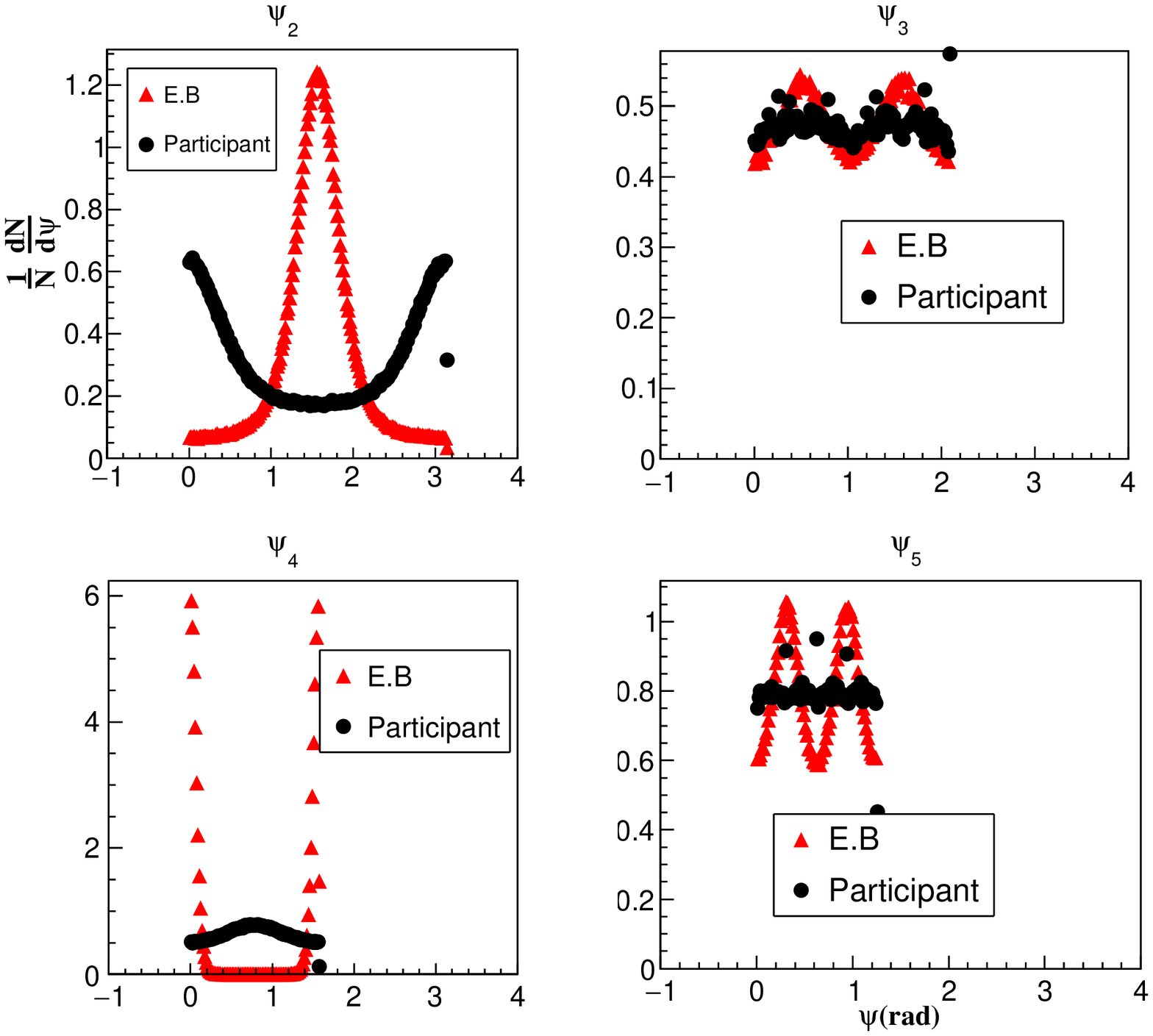}
\caption{(Color online) Top panel: distribution of $\psi_{\mathcal{E}}$ and $\psi_{P}$ for Ru+Ru collisions at $\sqrt{s_{NN}} = 200$ GeV for $b$=5 fm. Bottom panel: same as above but for impact parameter b=10fm.}
\label{fig:corrPsiEBRub5}
\end{figure}

\begin{figure}[htb]
\centering
\includegraphics[width=.40\textwidth]{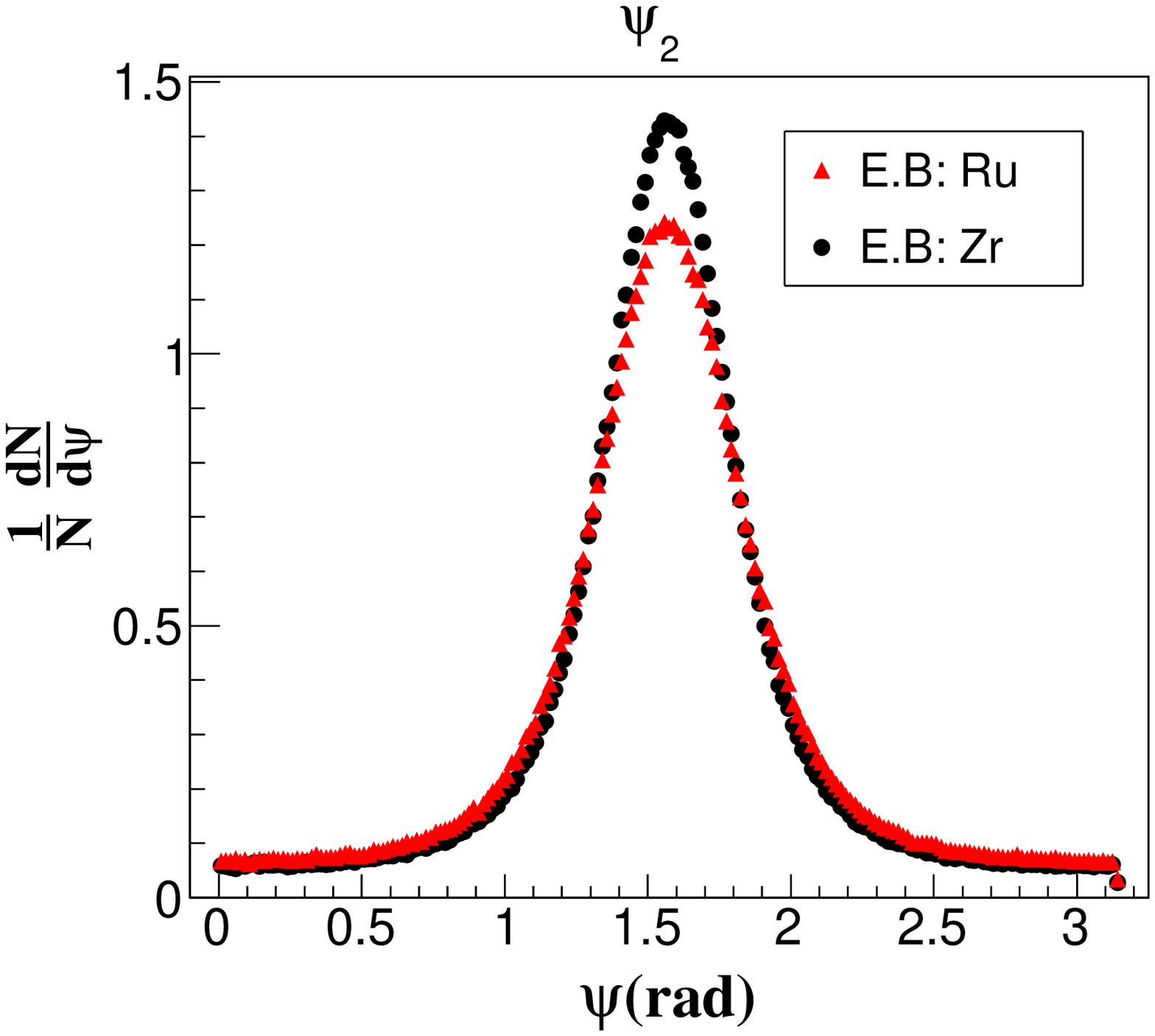}
\caption{(Color online) distribution of the second order plane $\psi_{\mathcal{E}}^{2}$ for Ru+Ru and Zr+Zr collisions at $\sqrt{s_{NN}} = 200$ GeV for $b$=10 fm.}
\label{fig:corrPsi2RuZrb10}
\end{figure}

We checked that for $b=0$ Ru+Ru, and Zr+Zr collisions (not shown here), results are very similar to what was observed for Au+Au collisions.
In the top panel of Fig.\ref{fig:corrPsiEBRub5}, we show $\psi_{\mathcal{E}}^{i}$ and $\psi_{P}^{i}$ ($i=2-5$ ) distribution for  
Ru+Ru collisions at $\sqrt{s_{NN}} = 200$ GeV for $b$=5 fm. The bottom panel of the same figure corresponds to results for $b$=10 fm collisions. 
A similar result was obtained (not shown here) for the Zr+Zr collisions.
It is interesting to note that $\psi_{\mathcal{E}}$'s and $\psi_{P}$'s in these small collision systems show a similar correlation as was observed for the Au+Au collisions. Like peripheral Au+Au collisions, we also observe the rotation of $\psi_{\mathcal{E}}$ by $\pi/2$ for the peripheral Ru+Ru and Zr+Zr collisions. Because the electromagnetic field produced in Ru+Ru and Zr+Zr collisions differs, we compared $\psi_{\mathcal{E}}^{2}$
distribution for $b=0$ fm. This fact is shown in Fig.\ref{fig:corrPsi2RuZrb10}, the field is higher for Zr+Zr compared to Ru+Ru; possibly as a consequence, we observe a slightly narrow peak for the Zr+Zr (black dots).  

\section{Conclusion and summary}
\label{Section: summary}

In summary, we have studied the event-by-event fluctuations of the electric and the magnetic fields and their possible correlation
with the geometry of the high-energy heavy-ion collisions. More particularly we studied the distribution of $\bf{E}\cdot\bf{B}$($=\mathcal{E}$)
in the transverse plane for Au+Au, Ru+Ru, and Zr+Zr collisions at $\sqrt{s_{NN}}=200$ GeV. Further, we show the $\tau$ and $\eta$ dependence
of $\mathcal{E}$ in Au+Au at 200 GeV per nucleon collisions. As expected, $\mathcal{E}$ is found to be symmetric in $\eta$ (around $\eta=0$),
and $\mathcal{E}$ quickly decays as a function of $\tau$ at a given $\eta$. Because $\mathcal{E}$ may contribute 
to CME as a source of the anomalous current, we investigate the centrality (impact parameter) dependence of the symmetry plane angle 
$\psi_{\mathcal{E}}$ and its possible correlation with the participant plane. We show that $\psi_{\mathcal{E}}$ is strongly correlated with 
$\psi_{P}$ for second-fifth order harmonics for Au+Au, Ru+Ru, and Zr+Zr collisions.  The second-order planes $\psi_{\mathcal{E}}$ and $\psi_{P}$ are not only correlated but also mostly coincides with each other except for the peripheral collisions, where a rotation by $\pi/2$ is observed for $\psi_{\mathcal{E}}$ irrespective of the collision system size. This phenomenon seems to be happening due to the almost cancellation of electric fields and dominating magnetic field pointing perpendicular to the participant plane in peripheral collisions. 
To conclude, in this exploratory study, we show that, like the magnetic fields, $\mathcal{E}$ is also correlated to the geometry of the collision even when we consider e-by-e fluctuation of nucleon positions.


\section*{Acknowledgements}
\label{Section:Acknowledgements}

We are thankful to the grid computing facility at Variable Energy Cyclotron Centre, Kolkata, for providing us CPU time. VR acknowledges financial support from the DST Inspire faculty research grant (IFA-16-PH-167), India.

\end{document}